\documentclass[aps,prd,superscriptadress,preprint]{revtex4}
\usepackage{graphicx}
\usepackage{amsmath}
\usepackage{amssymb}
\usepackage{booktabs}
\usepackage{tabularx}
\usepackage{textcomp}
 \usepackage{color}

\begin{document}
\title{Nonlinear susceptibilities under the framework of Dyson-Schwinger equations}

\author{A-Meng Zhao$^{1,}$}~\email[]{Email:zhaoameng@cxxy.seu.edu.cn}
\author{Zhu-Fang Cui$^{2,5}$}
\author{Yu Jiang$^{3,5}$}
\author{Hong-Shi Zong$^{2,4,5,}$}~\email[]{Email:zonghs@nju.edu.cn}
\address{$^{1}$ Department of Foundation, Southeast University Chengxian College, Nanjing 210088, China}
\address{$^{2}$ Department of Physics, Nanjing University, Nanjing 210093, China}
\address{$^{3}$ Center for Statistical and Theoretical Condensed Matter Physics, Zhejiang Normal University, Jinhua City, Zhejiang Province 321004, China}
\address{$^{4}$ Joint Center for Particle, Nuclear Physics and Cosmology, Nanjing 210093, China}
\address{$^{5}$ State Key Laboratory of Theoretical Physics, Institute of Theoretical Physics, CAS, Beijing 100190, China}

\begin{abstract}
Since the baryon-number susceptibilities are correlated with the cumulant of baryon-number fluctuations in experiments, we do calculations of the susceptibilities and compare them with the experimental fluctuation data under the framework of Dyson-Schwinger equations (DSEs) approach.  We compare our results with lattice QCD and experimental data at RHIC. The fitness of the results indicates that under the framework of DSEs, we can deal with the problems of heavy ion collisions properly.


\bigskip

\noindent PACS Number(s) 12.38.Lg, 11.10.Wx, 12.38.Mh, 25.75.Nq

\end{abstract}

\maketitle

\section{Introduction }
For a long time people believe that quark-number (or baryon-number) susceptibility (the second order) should develop some singularity \cite{a14,a15} near the critical end point (CEP) \cite{a16} of the quantum chromodynamics (QCD) phase transitions \cite{a17,a18} from hadronic matter to the quark-gluon plasma (QGP). To determine the location of CEP, a lot of phenomenological models \cite{a19,a20,a21,a22,a23,a24,c1,c3,c2} and lattice QCD \cite{a25,a26,a27} calculations are carried out.

It is well known that the $n$th cumulant of baryon-number fluctuations is proportional to the $n$th order of baryon-number susceptibilities \cite{nxu,a6,a28}. The baryon-number fluctuations, especially the variance, the skewness and the kurtosis, are experimental observables (In this paper, the experimental data comes from the STAR experiment at RHIC \cite{a8}). When studying the quark numbers at finite chemical potential by the fundamental theories of QCD, it is found that the quark-number density is determined by the corresponding dressed quark propagator only \cite{a1}. Then by generalizing this conclusion to the most universal situation of finite temperature and chemical potential \cite{a2}, we can calculate the $n$th order susceptibilities at finite temperature and chemical potential, and compare them with the experimental data from RHIC. Here, the crucial factor of getting a reasonable result from the susceptibilities is to adopt a reliable dressed quark propagator at finite temperature and chemical potential. In this paper, we obtain the dressed quark propagator under the framework of Dyson-Schwinger equations (DSEs) approach. The advantage of DSEs approach \cite{a29,a30,a31} is to provide a nonperturbative method to deal with dynamical chiral symmetry breaking and confinement at the same time. Therefore it is thought to be suitable to explore the QCD phase transition from hadronic matter to quark-gluon plasma \cite{a32}.

\section{Nonlinear susceptibilities in the DSEs framework}
\label{two}
From the first principle of QCD theory at zero temperature and finite chemical potential, the quark-number density is determined by the dressed quark propagator at finite chemical potential only \cite{a1},
\begin{equation}
\rho(\mu)=(-)N_cN_fZ_2\int\frac{d^4p}{(2\pi)^4}tr_{\gamma}[G[\mu](p)\gamma_4],
\end{equation}
where $N_c$ and $N_f$ represent the number of colors and flavors, respectively, and $G[\mu](p)$ is the quark propagator; furthermore, under the rainbow approximation of the Dyson-Schwinger equations, if we ignore the $\mu$ dependence of the dressed gluon propagator and assume that the dressed quark propagator at finite $\mu$ is analytic in the neighborhood of $\mu=0$, then we can obtain the following expression \cite{c4,c5}
\begin{equation}
G^{-1}[\mu](p)=G^{-1}(\widetilde{p})
\end{equation}
where~$\widetilde{p}=(\vec{p},p_4+i\mu)$, $\mu$ is the quark chemical potential, $Z_2=Z_2(\zeta^2,\Lambda^2)$ is the quark wave-function renormalization constant ($\zeta$ is the renormalization point and $\Lambda$ is the regularization mass-scale).

By replacing the integration over the fourth component of momentum with explicit summation over Matsubara frequencies, then this conclusion is generalized to the situation at finite temperature \cite{a2},
\begin{equation}
\rho(T,\mu)=(-)N_cN_fT\sum_{i=-\infty}^{+\infty}\int\frac{d^3p}{(2\pi)^3}tr_{\gamma}[G(\widetilde{p}_n)\gamma_4],
\end{equation}
where $p_4=\omega_n+i\mu$ with fermion frequencies $\omega_n=(2n+1)\pi T$, and in this study we put the regularization mass scale at infinity so that all renormalization constants including $Z_2$ are $1$.

The relation between the baryon-number density and the quark-number density is that, $\rho_B=\frac{1}{3}\rho(T,\mu)$. Then the ($n$-1)-th derivatives of $\rho_B$, by the baryon chemical potential $\mu_B$, are defined as the nonlinear susceptibilities of baryons of order $n$ \cite{a3}.
\begin{equation}
\chi^{(n)}_B=\frac{\partial ^{n-1}}{\partial \mu^{n-1}_B}\rho_B
=\frac{\partial ^{n-1}}{3^n\partial \mu^{n-1}}\rho(T,\mu)
=(-)\frac{N_cN_fT}{3^n}\sum_{i=-\infty}^{+\infty}\int\frac{d^3p}{(2\pi)^3}tr_{\gamma}[\frac{\partial ^{n-1}G(\widetilde{p}_n)}{\partial \mu^{n-1}}\gamma_4].
\label{chi}
\end{equation}

In dealing with the derivatives of the dressed quark propagator, we adopt the following identity
\begin{equation}
\frac{\partial G(\widetilde{p}_n)}{\partial \mu}=-G(\widetilde{p}_n)\frac{\partial G^{-1}(\widetilde{p}_n)}{\partial \mu}G(\widetilde{p}_n).
\end{equation}

According to the Ward identity, we can get the expression \cite{a2}
\begin{equation}
\Gamma^{(1)}_4(\widetilde{p}_n,0)=-\frac{\partial G^{-1}(\widetilde{p}_n)}{\partial \mu},
\end{equation}
then
\begin{equation}
\frac{\partial G(\widetilde{p}_n)}{\partial \mu}=G(\widetilde{p}_n)\Gamma^{(1)}_4(\widetilde{p}_n,0)G(\widetilde{p}_n).
\label{g1}
\end{equation}
Similarly, we get the following expressions
\begin{equation}
\frac{\partial^2 G(\widetilde{p}_n)}{\partial \mu^2}=G(\widetilde{p}_n)[2\Gamma^{(1)}_4(\widetilde{p}_n,0)G(\widetilde{p}_n)\Gamma^{(1)}_4(\widetilde{p}_n,0)+\Gamma^{(2)}_4(\widetilde{p}_n,0)]G(\widetilde{p}_n),
\label{g2}
\end{equation}
\begin{equation}
\begin{split}
\frac{\partial^3 G(\widetilde{p}_n)}{\partial \mu^3}=&G(\widetilde{p}_n)[6\Gamma^{(1)}_4(\widetilde{p}_n,0)G(\widetilde{p}_n)\Gamma^{(1)}_4(\widetilde{p}_n,0)G(\widetilde{p}_n)\Gamma^{(1)}_4(\widetilde{p}_n,0)
+3\Gamma^{(1)}_4(\widetilde{p}_n,0)G(\widetilde{p}_n)\Gamma^{(2)}_4(\widetilde{p}_n,0)\\
&+3\Gamma^{(2)}_4(\widetilde{p}_n,0)G(\widetilde{p}_n)\Gamma^{(1)}_4(\widetilde{p}_n,0)+\Gamma^{(3)}_4(\widetilde{p}_n,0)]G(\widetilde{p}_n),
\end{split}
\label{g3}
\end{equation}
where
\begin{eqnarray}
\begin{split}
\Gamma^{(2)}_4(\widetilde{p}_n,0)=\frac{\Gamma^{(1)}_4(\widetilde{p}_n,0)}{\partial \mu}=-\frac{\partial^2 G^{-1}(\widetilde{p}_n)}{\partial \mu^2},\\
\Gamma^{(3)}_4(\widetilde{p}_n,0)=\frac{\Gamma^{(2)}_4(\widetilde{p}_n,0)}{\partial \mu}=-\frac{\partial^3 G^{-1}(\widetilde{p}_n)}{\partial \mu^3}.\\
\end{split}
\end{eqnarray}

In order to get a reasonable dressed quark propagator at finite temperature and chemical potential, we turn to the rainbow approximation of the Dyson-Schwinger equations, as mentioned in Ref. \cite{a2}
\begin{equation}
G(\widetilde{p}_k)^{-1}=i\gamma\cdot \widetilde{p}_k+m+\frac{4}{3}T\sum_{i=-\infty}^{+\infty}\int\frac{d^3q}{(2\pi)^3}g^2D^{eff}_{\mu\nu}(\widetilde{p}_k-\widetilde{q}_n)\gamma_{\mu}G(\widetilde{q}_n)\gamma_{\nu},
\end{equation}
and here we adopt the rank-1 separable model, in which the gluon propagator is proposed in Refs. \cite{a4,a5} as
\begin{equation}
g^2D^{eff}_{\mu\nu}(\widetilde{p}_k-\widetilde{q}_n)=\delta_{\mu\nu}D_0f_0(\widetilde{p}^{2}_k)f_0(\widetilde{q}^{2}_n),
\end{equation}
where $f_0(\widetilde{p}^{2}_n)=exp(-\widetilde{p}^{2}_n/\Lambda ^2)$, with $\Lambda=0.678GeV$, $D_0\Lambda^2=128.0$, and the degenerated light quark mass $m=6.6 MeV$ \cite{a5}, these parameters are found to be successful in describing light flavor pseudoscalar and vector meson observables.

At the same time, the quark propagator is generally decomposed as
\begin{equation}
G^{-1}(\widetilde{p}_k)=i\vec{\gamma}\cdot\vec{p}A(\widetilde{p}^{2}_k)+i\gamma_4\widetilde{\omega}_k C(\widetilde{p}^{2}_k)+B(\widetilde{p}^{2}_k).
\end{equation}

For the rank-1 separable model, the rainbow-DSEs solution is $A(\widetilde{p}^{2}_k)= C(\widetilde{p}^{2}_k)=1$ and $B(\widetilde{p}^{2}_k)=m+b(T,\mu)f_0(\widetilde{p}^{2}_k)$. Then the propagator is finally read as
\begin{equation}
G^{-1}(\widetilde{p}_k)=i\vec{\gamma}\cdot\vec{p}+i\gamma_4\widetilde{\omega}_k +m+b(T,\mu)f_0(\widetilde{p}^{2}_k).
\label{b}
\end{equation}

Following the expression above, we get the conclusion that

\begin{eqnarray}
\begin{split}
\frac{\partial G^{-1}(\widetilde{p}_n)}{\partial \mu}&=-\gamma_4+b^{(1)}(T,\mu)f_0(\widetilde{p}^{2}_k)+b(T,\mu)f^{(1)}_0(\widetilde{p}^{2}_k),\\
\frac{\partial^2 G^{-1}(\widetilde{p}_n)}{\partial \mu^2}&=b^{(2)}(T,\mu)f_0(\widetilde{p}^{2}_k)+2b^{(1)}(T,\mu)f^{(1)}_0(\widetilde{p}^{2}_k)+b(T,\mu)f^{(2)}_0(\widetilde{p}^{2}_k),\\
\frac{\partial^3 G^{-1}(\widetilde{p}_n)}{\partial \mu^3}&=b^{(3)}(T,\mu)f_0(\widetilde{p}^{2}_k)+3b^{(2)}(T,\mu)f^{(1)}_0(\widetilde{p}^{2}_k)+3b^{(1)}(T,\mu)f^{(2)}_0(\widetilde{p}^{2}_k)+b(T,\mu)f^{(3)}_0(\widetilde{p}^{2}_k),\\
\end{split}
\label{bn}
\end{eqnarray}
where $f^{(n)}_0(\widetilde{p}^{2}_k)$ is the $n$-th derivatives of $f_0(\widetilde{p}^{2}_k)$ by $\mu$, and $b^{(n)}(T,\mu)$ is the n-th derivatives of $b(T,\mu)$ by $\mu$ similarly. $b(T,\mu)$ and $b^{(n)}(T,\mu)$ are solved numerically, which are shown in the appendix in detail. Then substituting Eqs.~(\ref{g1}), (\ref{g2}), (\ref{g3}) and Eqs.~(\ref{b}), (\ref{bn}) into Eq.~(\ref{chi}), we can get the results of $\chi_B^{(n)}$.

\section{Results}
\label{three}

Our interest in the nonlinear susceptibilities comes from that they are related to cumulants of the baryon number fluctuations in a grand canonical ensemble \cite{a6,a7}. And the details of the correlation are \cite{nxu}:
\begin{eqnarray}
\begin{split}
S\sigma=\frac{T\chi^{(3)}_B}{\chi^{(2)}_B},\\
\kappa\sigma^2=\frac{T^2\chi^{(4)}_B}{\chi^{(2)}_B},\\
\frac{\kappa\sigma}{S}=\frac{T\chi^{(4)}_B}{\chi^{(3)}_B},\\
\end{split}
\end{eqnarray}
where $\sigma^2$ is the variance, $S$ is the skewness and $\kappa$ is the kurtosis.

\begin{figure}
\centering
\includegraphics[width=.7\linewidth]{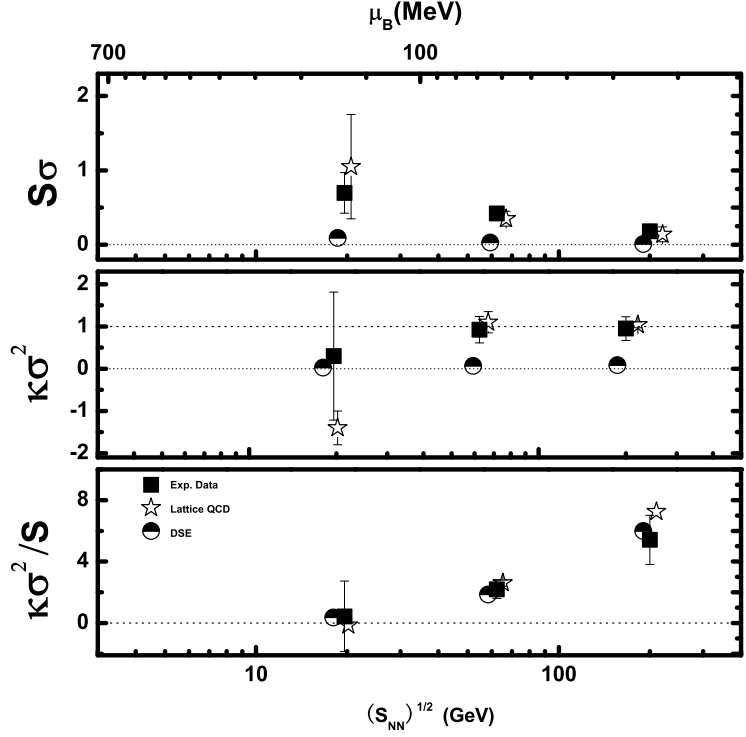}
\caption{Comparison of DSEs result, lattice QCD and experimental data for $S\sigma$, $\kappa\sigma^2$ and $\frac{\kappa\sigma}{S}$ at $\sqrt{S_{NN}}=19.6, 62.4, 200$ GeV. The black boxes are the experimental data and the stars are the lattice results. Our DSEs results are shown by the circles.}
\label{ex}
\end{figure}

In Fig. \ref{ex}, $S\sigma$, $\kappa\sigma^2$ and $\frac{\kappa\sigma}{S}$ are shown as a function of $\sqrt{S_{NN}}$ for $Au+Au$ collisions at RHIC. The corresponding freeze-out chemical potential $\mu_B$ to $\sqrt{S_{NN}}$ is also shown on the top of the picture. The correlations between $\sqrt{S_{NN}}$ and the bulk properties ($\mu_B$ and T) of chemical freeze-out are discussed in Refs. \cite{a9,a10,a12,a13}. Here we adopt that
\begin{eqnarray}
\begin{split}
T(\mu_B)=a-b\mu^2_B-c\mu^4_B,\\
\mu_B(\sqrt{S_{NN}})=\frac{d}{1+e\sqrt{S_{NN}}},\\
\end{split}
\label{24}
\end{eqnarray}
where $a=0.166\pm0.002 GeV$, $b=0.139\pm0.016 {GeV}^{-1}$, $c=0.053\pm0.021 {GeV}^{-3}$, $d=1.308\pm0.028 GeV$ and $e=0.273\pm0.008 {GeV}^{-1}$ \cite{a12}. In Tab. \ref{tab1}, the corresponding $T$, $\mu_B$ and $\mu$ to $\sqrt{S_{NN}}=19.6, 62.4, 200$ GeV are calculated by Eq. (\ref{24}) respectively.
\begin{table}[htbp]
 \caption{\label{tab1}Correlation between $\sqrt{S_{NN}}$ , temperature , baryon and quark chemical potential.}
 \begin{tabular}{|c|c|c|c|}
 \hline
  $\sqrt{S_{NN}}$(GeV)    & T(MeV)    & $\mu_B$(MeV)&$\mu$(MeV)\\
  \hline
  19.6 & 159 & 229 &77\\
  \hline
   62.4 & 165 & 82 &28\\
  \hline
   200 & 166 & 27 &9\\
  \hline
 \end{tabular}
\end{table}
The results that we obtain under the framework of Dyson-Schwinger equations are compared with lattice QCD and experimental data. The lattice QCD calculations, with a cutoff of $1/a\cong960$ to $1000MeV$, was carried out by using two flavors of quark \cite{a11}. The experimental data comes from $Au+Au$ collisions at RHIC, in which impact parameter values are less than $3fm$ \cite{a8}.

\begin{figure}
\centering
\includegraphics[width=.7\linewidth]{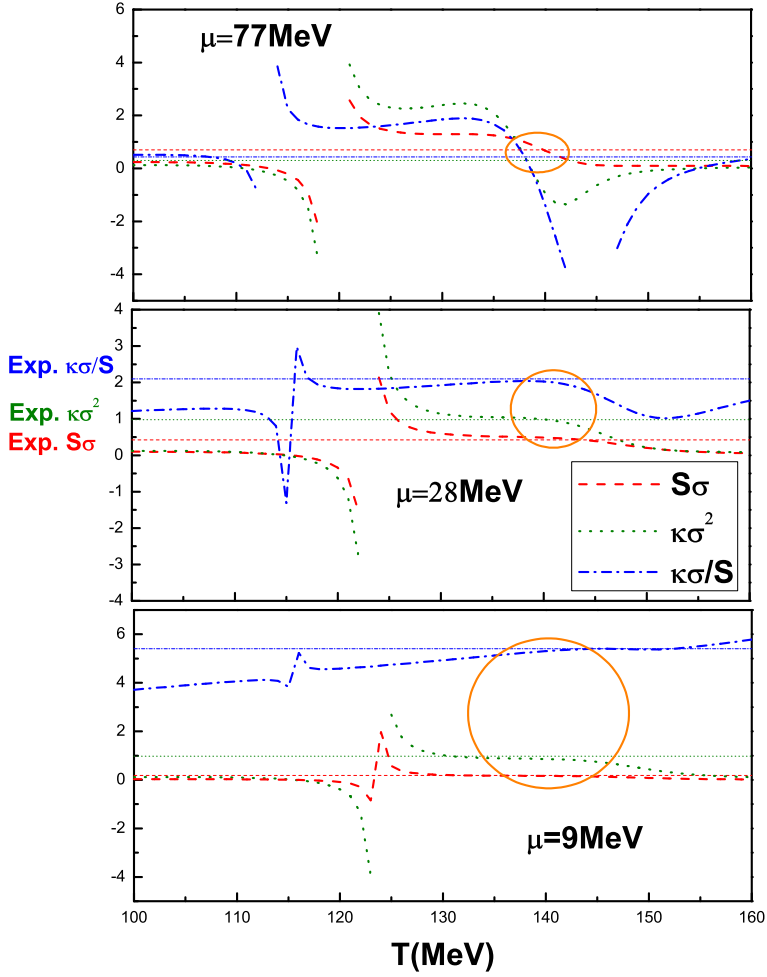}
\caption{(Color online) The results of $S\sigma$, $\kappa\sigma^2$ and $\frac{\kappa\sigma}{S}$ are shown in three curves as a function of $T$ in each plot. The three horizontal lines in each plot represent the corresponding experimental results of $S\sigma$, $\kappa\sigma^2$ and $\frac{\kappa\sigma}{S}$ at RHIC.  And the circle in each plot demonstrates the region where the three experimental data (horizontal lines) all have a intersection with the corresponding DSEs result curves.}
\label{140}
\end{figure}

In Fig. \ref{ex}, it is shown that, comparing with the lattice data, our DSEs results demonstrate less fitness with the experimental data on the top two plots. Conversely, as to the value of $\frac{\kappa\sigma}{S}$, our results fit better to the experimental data than the lattice. To explore what makes this difference, then we fix the value of quark chemical potential $\mu$ at $77, 28$ and $9$ MeV, and calculate $S\sigma$, $\kappa\sigma^2$ and $\frac{\kappa\sigma}{S}$ by changing the temperature $T$ from $100$ to $160$ MeV. And the results are shown in Fig.~\ref{140}.

The motivation of our exploration of the temperature region smaller than $160$ MeV comes from the conclusion obtained in Ref.~\cite{a2}, which adopt the similar approximations as ours. In that paper, it is concluded that while the chemical potential of the CEP obtaining from the rank-1 separable model ($\mu_{CEP}=164$ MeV) is located in the region of the experimental estimate($\mu_{CEP}\sim 150-180$ MeV)~\cite{c6}, which is obtained by extracting $\eta/s$ from an elliptic flow excitation function, the CEP temperature $T_{CEP}=117$ MeV  is smaller than its corresponding experimental estimated results $T_{CEP}\sim 165-170$ MeV~\cite{c6}. Besides, Ref.~\cite{a2} gives the pseudo-critical temperature  at $T_c=150$ MeV , it is also smaller than the value of $T_c=175$ MeV~\cite{nxu}, which is obtained through a comparison of thermodynamic fluctuations predicted in lattice with the experimental data. This characteristic of $\mu$ and $T$ inspires us to study the temperatures smaller than those in Tab.~\ref{tab1} and at the same time fixed the chemical potentials unchanged as in Tab.~\ref{tab1}.

In Fig.~\ref{140} the results of $S\sigma$, $\kappa\sigma^2$ and $\frac{\kappa\sigma}{S}$(three curves) are shown as a function of $T$ at $\mu=77, 28$ and $9$ MeV respectively. The three horizontal lines in each plot of Fig. \ref{140} represent the corresponding results of $S\sigma$, $\kappa\sigma^2$ and $\frac{\kappa\sigma}{S}$ at RHIC experiments. Inside the region of the circle in each plot, the three experimental data(horizontal lines) all have a intersection with the corresponding DSEs result curves. That is to say, in this region the DSEs results fit the experimental data completely. And the region, shown in the three plots, are during the value of $T$ from $138$ to $145$ MeV, which is approximately $20$ MeV smaller than the temperature given in Tab. \ref{tab1}. The reduction of temperature is consistent with the conclusion of Ref.~\cite{a2}. Combining the results obtaining in Ref.~\cite{a2} and this paper, it indicates that if we adopt the rank-1 separable model of DSEs, Eq.~(\ref{24}) is not suitable to determine the freeze-out temperature correlated to a certain $\sqrt{S_{NN}}$.

\begin{figure}
\centering
\includegraphics[width=.7\linewidth]{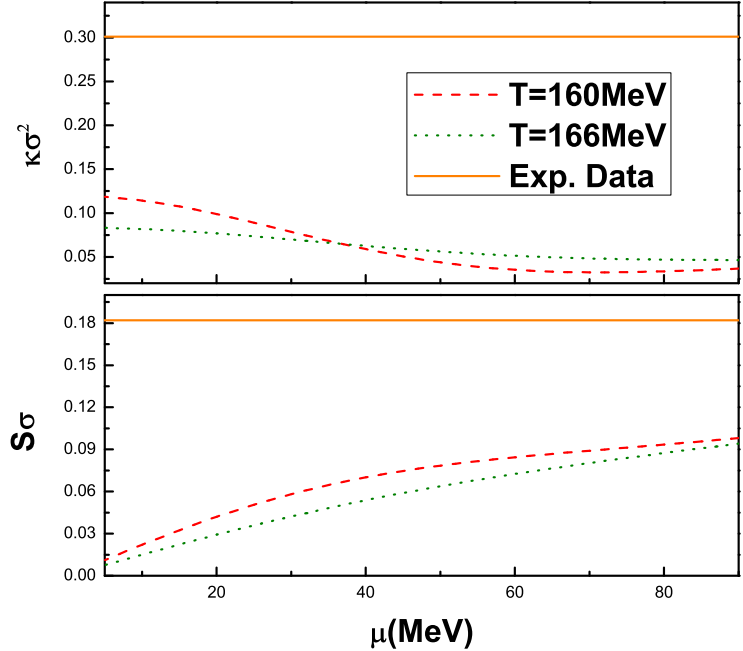}
\caption{(Color online) $S\sigma$ and $\kappa\sigma^2$ are shown as a function of $\mu$ in each plot. And the horizontal line represents the smallest one of the corresponding experimental data at $\sqrt{S_{NN}}=19.6, 62.4$ and $200$ GeV. }
\label{mix}
\end{figure}

Finally as a supplement to Fig. \ref{140}, we fix $T$ at $160$ and $166$ MeV, and calculate $S\sigma$, $\kappa\sigma^2$ and $\frac{\kappa\sigma}{S}$ as a function of $\mu$. The results of $S\sigma$ and $\kappa\sigma^2$ are shown in Fig. \ref{mix}. The curves in each plot represent our results and the horizontal line represents the smallest one of the corresponding experimental data at $\sqrt{S_{NN}}=19.6, 62.4$ and $200$ GeV. In Fig. \ref{mix}, we can see that the DSEs results are all too small to compare with the experimental data. It indicates once more that while the chemical potential determined by Eq.~(\ref{24}) is acceptable, the temperature value determined is inadequate for the rank-1 separable model of DSEs. That the reason caused this $T$ reduction comes from whether our simplification of the dressed quark propagator or the approximations of the rank-1 separable model is one aspect of our further study, since the rank-1 separable model is in some sense a big approximation of the gluon propagator. Actually, in order to draw some more reliable conclusions, some further studies of us by adopting more elegant gluon models (such as Refs.~\cite{c3,c2}) is on the road~\cite{cnew}.

According to the three figures above, it can be concluded that: following the values of $T$ and $\mu$ shown in Tab. \ref{tab1}, which is determined by Eq. (\ref{24}), our results of $S\sigma$ and $\kappa\sigma^2$ can not fit the experimental data well; if we fix $\mu$ at $77, 28$ and $9$ MeV and show $S\sigma$, $\kappa\sigma^2$ and $\frac{\kappa\sigma}{S}$ as a function of $T$, it is found that the DSEs results fit the experimental results completely well when $T$ changes from $138$ to $145$ MeV; no matter what value of $\mu$ is chosen, our results are much smaller than the experimental data during the region of temperature given by Tab.~\ref{tab1}~(from $159$ to $166$ MeV).

\section{Summary}
\label{four}
Except the singularity near the critical end point(CEP) of QCD phase transition, the baryon-number susceptibilities are also correlated with the cumulant of baryon-number fluctuations. Therefore we can get the fluctuations by the calculation of the susceptibilities and compare them with the experimental data. According to the QCD theories the quark-number density is determined by the dressed quark propagator only. Since then the problem is covert to find a reliable dressed quark propagator. Here we adopt the dressed quark propagator under  the framework of DSEs approach, which is thought to be suitable to study the QCD phase transitions.

We compare our results obtained under the DSEs framework with lattice QCD and experimental data at RHIC. The DSEs results can fit the experimental data of $\frac{\kappa\sigma}{S}$ well. But if the region of $T$ is moved from $159-166$ MeV to $138-145$ MeV, the DSEs results fit the experimental data completely well. The fitness indicates that the method of Dyson-Schwinger equations is reliable and productive in dealing with the relativistic heavy ion collisions. To solve the Dyson-Schwinger equations, we adopt the rank-1 separable model which makes a simplification to the gluon propagator. Actually, in order to draw some more reliable conclusions, some further studies of us by adopting more elegant gluon models (such as Refs.~\cite{c3,c2}) is already on the road. At the same time, we do not take into account the influence of the magnetic field that probably created in QGP. And these two aspects are the directions for our further studies.

\section*{Acknowledgements}
This work is supported in part by the National Natural Science Foundation of China (under Grants No. 11275097, No. 11475085, and No. 11105122), the National Basic Research Program of China (under Grant No. 2012CB921504), and the Jiangsu Planned Projects for Postdoctoral Research Funds (under Grant No. 1402006C).

\appendix
\section{$\mathbf{b(T,\mu)}$ and its $n$-th derivatives}\label{appA}
For the rank-1 separable model, the rainbow-DSEs solution is $A(\widetilde{p}^{2}_k)= C(\widetilde{p}^{2}_k)=1$ and $B(\widetilde{p}^{2}_k)=m+b(T,\mu)f_0(\widetilde{p}^{2}_k)$, where $b(T,\mu)$ satisfies the following equation

\begin{equation}
b(T,\mu)=\frac{16}{3}D_0T\sum_{i=-\infty}^{+\infty}\int\frac{d^3q}{(2\pi)^3}\frac{f_0(\widetilde{q}^{2}_n)[m+b(T,\mu)f_0(\widetilde{q}^{2}_n)]}{[\widetilde{q}^{2}_n+(m+b(T,\mu)f_0(\widetilde{q}^{2}_n))^2]}.
\label{b1}
\end{equation}
we can obtain the value of $b(T,\mu)$ by solving Eq.~(\ref{b1}) numerically.

We define that
\begin{eqnarray}
\begin{split}
w_1&=f_0(\widetilde{q}^{2}_n)[m+b(T,\mu)f_0(\widetilde{q}^{2}_n)],\\
w_2&=[\widetilde{q}^{2}_n+(m+b(T,\mu)f_0(\widetilde{q}^{2}_n))^2],\\
\end{split}
\end{eqnarray}
then by taking derivatives of Eq.~(\ref{b1}), we get a new equation of $b^{(1)}(\mu,T)$

\begin{equation}
b^{(1)}(\mu,T)=\frac{\partial b(T,\mu) }{\partial \mu}=\frac{16}{3}D_0T\sum_{i=-\infty}^{+\infty}\int\frac{d^3q}{(2\pi)^3}\frac{w^{(1)}_1w_2-w_1w^{(1)}_2}{w^2_2},
\label{b2}
\end{equation}
where
\begin{eqnarray}
\begin{split}
w^{(1)}_1&=\frac{\partial w_1 }{\partial \mu}=f^{(1)}_0(\widetilde{q}^{2}_n)(m+b(T,\mu)f_0(\widetilde{q}^{2}_n))+f_0(\widetilde{q}^{2}_n)(b(T,\mu)f^{(1)}_0(\widetilde{q}^{2}_n)+b^{(1)}(T,\mu)f_0(\widetilde{q}^{2}_n)),\\
w^{(1)}_2&=\frac{\partial w_2 }{\partial \mu}=2(m+b(T,\mu)f_0(\widetilde{q}^{2}_n))(b(T,\mu)f^{(1)}_0(\widetilde{q}^{2}_n)+b^{(1)}(T,\mu)f_0(\widetilde{q}^{2}_n))+2i\widetilde{\omega}_n,\\
f^{(1)}_0(\widetilde{q}^{2}_n)&=\frac{\partial}{\partial\mu}f_0(\widetilde{q}^{2}_n)=-\frac{2i\widetilde{\omega}_n}{\Lambda^2}f_0(\widetilde{q}^{2}_n),
\end{split}
\end{eqnarray}
then similarly, we can get $b^{(1)}(\mu,T)$ by solving Eq.\ref{b2}.

Then by the same way, we obtain the second and third derivatives of  $b(T,\mu)$:
\begin{equation}
b^{(2)}(\mu,T)=\frac{\partial^2 b(T,\mu) }{\partial \mu^2}=\frac{16}{3}D_0T\sum_{i=-\infty}^{+\infty}\int\frac{d^3q}{(2\pi)^3}\frac{(w^{(2)}_1w_2-w_1w^{(2)}_2)w_2-2w^{(1)}_2(w^{(1)}_1w_2-w_1w^{(1)}_2)}{w^3_2},
\end{equation}

where
\begin{eqnarray}
\begin{split}
w^{(2)}_1=&f^{(2)}_0(\widetilde{q}^{2}_n)(m+b(T,\mu)f_0(\widetilde{q}^{2}_n))+2f^{(1)}_0(\widetilde{q}^{2}_n)(b(T,\mu)f^{(1)}_0+b^{(1)}(T,\mu)f_0(\widetilde{q}^{2}_n))\\
&+f_0(\widetilde{q}^{2}_n)(b(T,\mu)f^{(2)}_0(\widetilde{q}^{2}_n)+2b^{(2)}(T,\mu)f^{(1)}_0(\widetilde{q}^{2}_n)+b^{(2)}(T,\mu)f_0(\widetilde{q}^{2}_n)),\\
w^{(2)}_2=&-2+2(b(T,\mu)f^{(1)}_0(\widetilde{q}^{2}_n)+b^{(1)}(T,\mu)f_0(\widetilde{q}^{2}_n))^2\\
&+2(m+b(T,\mu)f_0(\widetilde{q}^{2}_n))(b(T,\mu)f^{(2)}_0(\widetilde{q}^{2}_n)+2b^{(1)}(T,\mu)f^{(1)}_0(\widetilde{q}^{2}_n)+b^{(2)}(T,\mu)f_0(\widetilde{q}^{2}_n))\\
f^{(2)}_0(\widetilde{q}^{2}_n)=&\frac{\partial^2}{\partial\mu^2}f_0(\widetilde{q}^{2}_n)=\frac{2}{\Lambda^2}f_0(\widetilde{q}^{2}_n)-\frac{2i\widetilde{\omega}_n}{\Lambda^2}f^{(1)}_0(\widetilde{q}^{2}_n),
\end{split}
\end{eqnarray}

and

\begin{eqnarray}
\begin{split}
b^{(3)}(\mu,T)=\frac{\partial^3 b(T,\mu) }{\partial \mu^3}=&\frac{16}{3}D_0T\sum_{i=-\infty}^{+\infty}\int\frac{d^3q}{(2\pi)^3}[\frac{w^{(3)}_1w_2-3w^{(2)}_1w^{(1)}_2-3w^{(1)}_1w^{(2)}_2-w_1w^{(3)}_2}{w^2_2}\\
&-6\frac{w_1w_2w^{(1)}_2w^{(2)}_2+w^{(1)}_1w_2(w^{(1)}_2)^2-w_1(w^{(1)}_2)^3}{w^4_2}],
\end{split}
\end{eqnarray}

where

\begin{eqnarray}
\begin{split}
w^{(3)}_1=&f^{(3)}_0(\widetilde{q}^{2}_n)(m+b(T,\mu)f_0(\widetilde{q}^{2}_n))+3f^{(2)}_0(\widetilde{q}^{2}_n)(b(T,\mu)f^{(1)}_0(\widetilde{q}^{2}_n)+b^{(1)}(T,\mu)f_0(\widetilde{q}^{2}_n))\\
&+3f^{(1)}_0(\widetilde{q}^{2}_n)(b(T,\mu)f^{(2)}_0(\widetilde{q}^{2}_n)+2b^{(1)}(T,\mu)f^{(1)}_0(\widetilde{q}^{2}_n)+b^{(2)}(T,\mu)f_0(\widetilde{q}^{2}_n))\\
&+f_0(\widetilde{q}^{2}_n)(b(T,\mu)f^{(3)}_0(\widetilde{q}^{2}_n)+3b^{(1)}(T,\mu)f^{(2)}_0(\widetilde{q}^{2}_n)+3b^{(2)}(T,\mu)f^{(1)}_0(\widetilde{q}^{2}_n)+b^{(3)}(T,\mu)f_0(\widetilde{q}^{2}_n)),\\
w^{(3)}_2=&6(b(T,\mu)f^{(1)}_0(\widetilde{q}^{2}_n)+b^{(1)}(T,\mu)f_0(\widetilde{q}^{2}_n))(b(T,\mu)f^{(2)}_0(\widetilde{q}^{2}_n)+2b^{(1)}(T,\mu)f^{(1)}_0(\widetilde{q}^{2}_n)\\
&+b^{(2)}(T,\mu)f_0(\widetilde{q}^{2}_n))+2(m+b(T,\mu)f_0(\widetilde{q}^{2}_n))\\
&*(b(T,\mu)f^{(3)}_0(\widetilde{q}^{2}_n)+3b^{(1)}(T,\mu)f^{(2)}_0(\widetilde{q}^{2}_n)+3b^{(2)}(T,\mu)f^{(1)}_0(\widetilde{q}^{2}_n)+b^{(3)}(T,\mu)f_0(\widetilde{q}^{2}_n)),\\
f^{(3)}_0(\widetilde{q}^{2}_n)=&\frac{\partial^3}{\partial\mu^3}f_0(\widetilde{q}^{2}_n)=\frac{4}{\Lambda^2}f^{(1)}_0(\widetilde{q}^{2}_n)-\frac{2i\widetilde{\omega}_n}{\Lambda^2}f^{(2)}_0(\widetilde{q}^{2}_n).
\end{split}
\end{eqnarray}

\end{document}